# Low Level Radio-Frequency system used for testing the RFQ prototype of MYRRHA *

C. Joly†, W. Sarlin, Y. Gargouri, J. Lesrel, J-F. Yaniche, S. Berthelot, A. Escalda, G. Ferry, B-Y. Ky, M. Pereira, Institut de Physique Nucléaire (UMR 8608) - CNRS/ IN2P3-Université Paris-Sud, Université Paris-Saclay, Orsay, France
J. Belmans, F. Davin, W. De Cock, P. Della Faille, F. Pompon, D. Vandeplassche
Centre d'études et de recherche Nucléaire- SCK•CEN, Louvain-La-Neuve, Belgium

*Abstract*
Within the framework of the European, project MYRTE (MYRRHA Research and Transmutation Endeavour) of the H2020 program, a 4-Rods RFQ (Radio Frequency Quadrupole) has been designed at 176.1 MHz RFQ for accelerating up to 4 mA protons in CW (Continuous Wave) operation from 30 keV up to 1.5 MeV. A LLRF prototype has been developed to regulate the amplitude and the phase of the accelerator field into the RFQ and the frequency of the RFQ controlling the motor of the frequency tuner. Here we present the facility at Louvain-la-Neuve, with a focus on its LLRF, as well as some preliminary results.

## INTRODUCTION

The MYRRHA (Multi-purpose Hybrid Research reactor for High-tech Applications) project mainly aims at realizing a hybrid reactor demonstrator, using an Accelerator Driven System (ADS) for transmuting radiotoxic waste [1, 2]. It is composed of a high CW power superconducting Linac (proton beam up to 2.4MW) associated to an external neutron flux source for feeding the sub-critical core of a nuclear reactor. The first phase of the MYRRHA project, called MINERVA (MYRRHA Isotopes productioN coupling the linEar acceleRator to the Versatile proton target fAcility), an Accelerator up to 100MeV and a Proton target Facility, was launched on the 7 September 2018 with the Belgian Government decision to allocate 556M€ for the period 2019-2038 for building the MYRRHA facility.

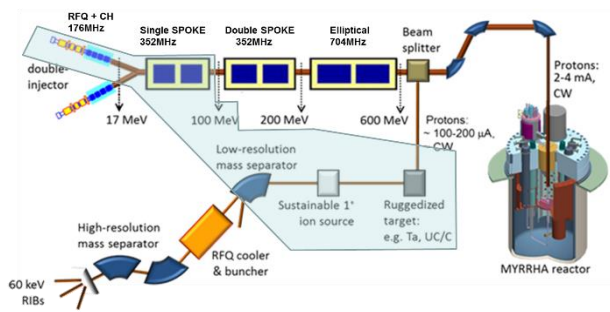

Figure 1: Conceptual layout of the MYRRHA Facility with in the cyan area, the MINERVA project (2019-2026).

Furthermore, several funding programmes have been performing for more than 20 years R&D to support the development of the MYRRHA facility and its ADS. Said R&D is addressing stringent requirements of our projects, such as high reliability with a MTBF > 250 hours, focusing on critical components such as the injector or the superconducting cavities. The main topic for the Accelerator R&D of the last European project called MYRTE is the Injector demonstration. Within this framework, a Low Level Radio Frequency prototype system for the 4-Rod Radio Frequency Quadrupole (RFQ) and a phase reference system (Master Oscillator) have been developed by IPNO, and installed into the SCK-CEN 5.9 MeV Injector test facility (UCL site) at Louvain-La-Neuve, Belgium.

## 5.9 MEV INJECTOR TEST FACILITY

The normal conducting injector up to 5.9 MeV [3] will be installed and tested into a bunker (21m x 4m) built by SCK-CEN within the Centre de Ressources du Cyclotron at UCL, Louvain-la-Neuve, Belgium.

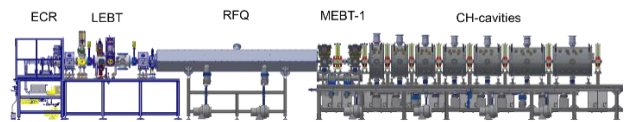

Figure 2: 5.9 MeV injector CAD scheme

Two deionized water circuits mainly dedicated to the RFQ, the CH cavities and RF amplifiers provide water cooling at 20°C ±0.5°C or in the 20-30° range as required by the system. 750 kVA in 3x400V input electrical power is available for the facility. Moreover, a vacuum system with six dry vacuum pumps, eleven turbomolecular vacuum pumps and seven Ion pumps managed by a Programmable Logic Controller (PLC) is being deployed.

To date, the ECR proton source from PANTECHNIK associated to low energy beam transport line developed by LPSC is operational [4].

The second part, the RFQ designed by IAP Frankfurt and built by NTG, was installed and connected to the cooling water and vacuum systems.

The 192kW@176.1MHz RF power solid state amplifier designed and built by IBA has been installed and characterized by the SCK•CEN and IBA teams [5] and it is

* This work is supported by the European Atomic Energy Community's (Euratom) H2020 Program under grant agreement n°662186186 (MYRTE Project)

used for the RFQ conditioning and the field regulation of the accelerator with the LLRF system.

## THE LOW LEVEL RF SYSTEM

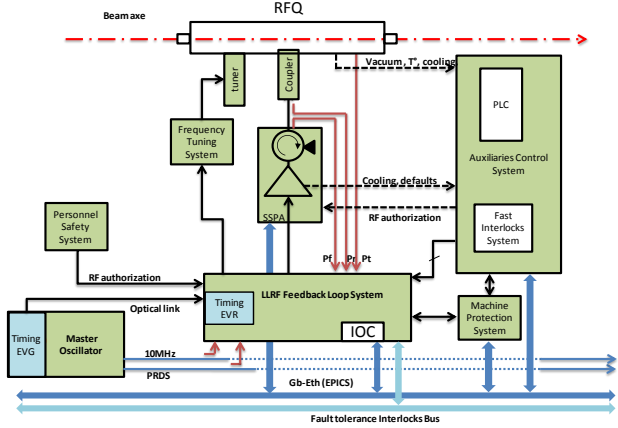

Figure 3: RFQ Control system.

Apart from the requirements linked with fault-tolerance aspects, the LLRF must ensure the amplitude and the phase stabilities of the accelerating field, ±0.2% and ±0.1° respectively.

An overview of the complete system is shown in Fig. 2 including the main functions:
- The Master Oscillator (MO), which provides all the reference signals needed by the LLRF systems, has been developed by IPNO. It integrates different components from the companies Wenzel associated and AR Electronique, France in order to obtain signals with very low phase noise until 704.4MHz.

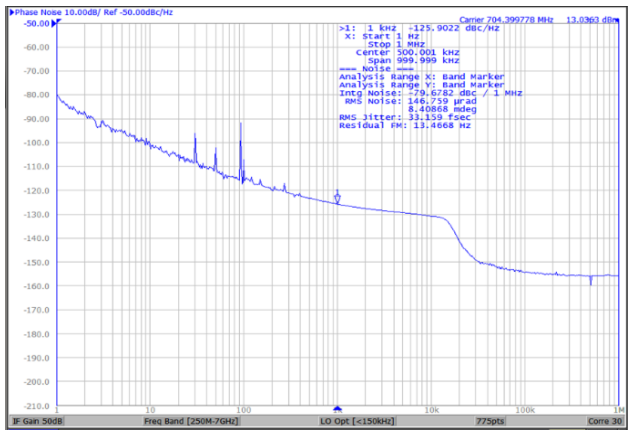

Figure 4: Phase noise measurements @704.4MHz with a 5052B analyser.

Table 1 presents the phase jitters results rms between 1Hz to 1MHz for the timing and accelerator references:

Table 1: Jitter measurements.

| Frequency | Jitter rms | Jitter rms |
|---|---|---|
| 88.05 MHz | 1.6 mDeg | 50 fs |
| 176.1 MHz | 2.0 mDeg | 33 fs |
| 352.2 MHz | 4.5 mDeg | 36 fs |
| 704.4 MHz | 8.4 mDeg | 33 fs |

- The LLRF feedback is composed of two main RF functions, an analogue Self-Excited Loop (SEL) operating at 176.1MHz including an actuator in the loop used in the generator driven mode and a RF front-end (4 inputs downconverter and one output vector modulator). The signal processing is implemented into an in-house digital motherboard called DALTON with two FMC slots and using a Virtex 6 Field-Programmable Gate Array (FPGA) linked to a Marvell Arm-5 processor by PCIe.

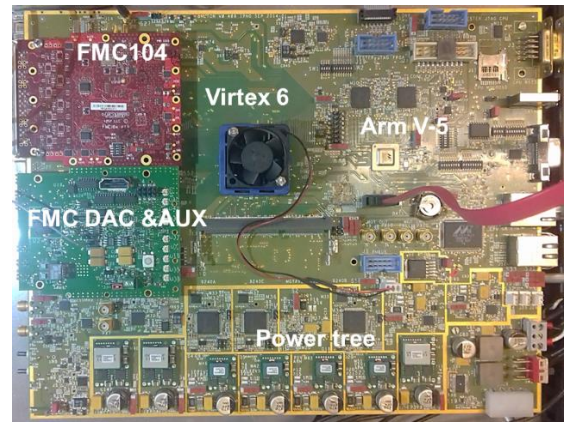

Figure 5: DALTON Board view with FMC mezzanine boards connected.

The FMC board from the 4DSP company referenced FMC104 samples the IF signals with four ADC, 14bit resolution@250MSPS max. The second FMC board, designed by IPNO, integrates several functions as the vector modulator I/Q commands by a two channels digital to analog converter, AD9747, 16bit@250MHz max. Another functions concerns the sampling of the slow signals, the motor controller interface and interlocks.

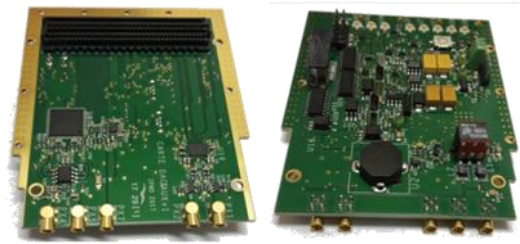

Figure 6: In-House FMC board called DAC&AUX

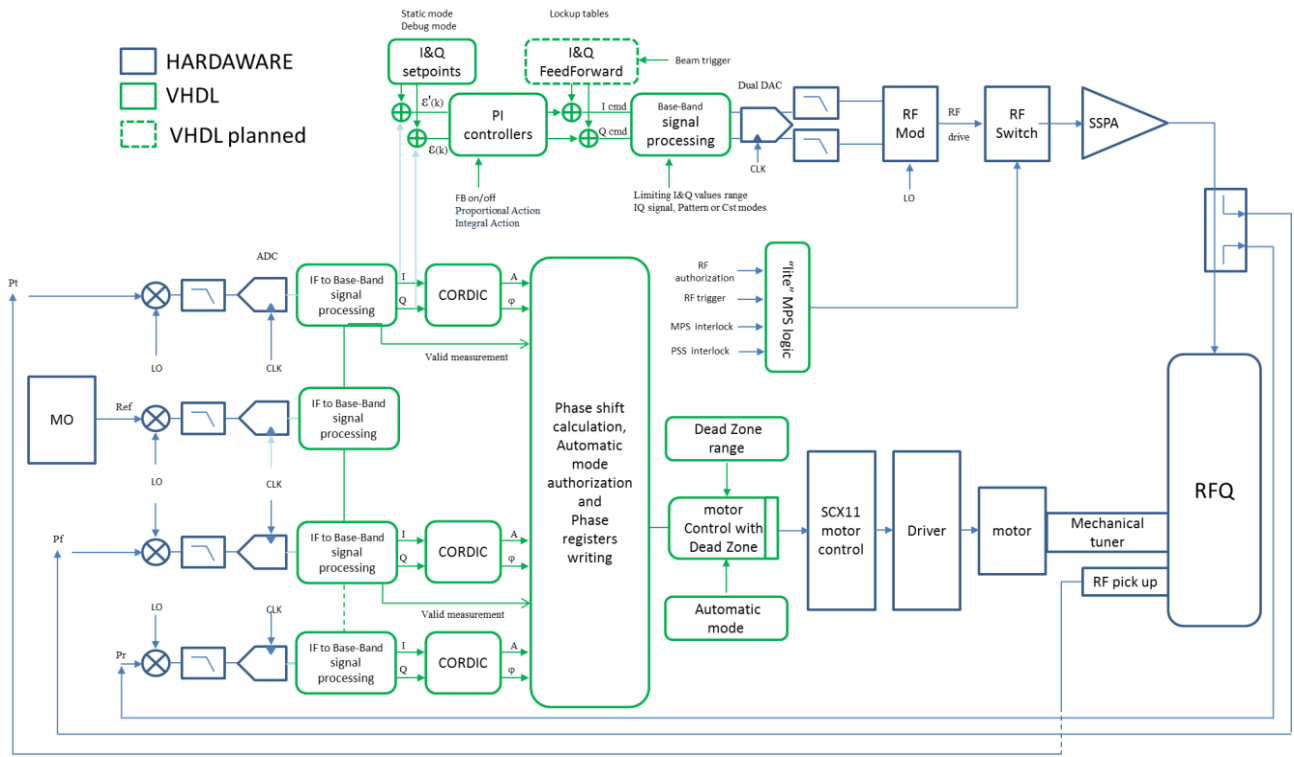

Figure 7: Main hardware and VHDL functions scheme.

- A Frequency Tuning System using a SCX11 controller associated to a power driver from ORIENTAL MOTOR moves a motorized plunger (100 mm range) for operating the RFQ at 176.1MHz, in function of the phase shift measured by the LLRF Feedback system.

- The PCI based timing event system, using the boards developed by the company Micro Research Finland Oy, has been deployed and integrated on the general control system using the EPICS (Experimental Physics and Industrial Control System) tools by the company Cosylab. It distributes in particular the RF and beam triggers for the LLRF system synchronization.

- The Personnel Safety System (PSS) is a classic system limiting access to the bunker using radioprotection measurements and non-authorized entrance detection to shutdown RF power by sending a PSS interlock to the LLRF system and the RF power amplifier.

- An independent Machine Protection System (MPS) is under development for the injector but some protections are already operational for testing the RFQ.

Associated to the hardware, VHDL code was implemented into the FPGA, as shown in Fig. 7. It a classic I/Q demodulation of the IF signals sampled at 40MHz after an offset correction is applied to the samples for correcting the ADC error. Then I/Q samples are filtered with FIR low pass filters and used for the error calculation needed to the Proportional Integral (PI) corrector. The IQ values in the output of the PI corrector can be limited in amplitude before driving the RF vector modulator. A second loop using the CORDIC IP from XILINX allows to calculate the phase shift between the forward and transmitted signals. In automatic mode, it is kept identical to the value corresponding to the 176.1MHz frequency tuning position, controlling the plunger motor via nine inputs and four outputs TTL signals from/to the SCX11 controller. A dead zone is used for restricting the movements.

## EPICS DEVELOPMENTS

To perform Control and Command (CC) of the LLRF system, an EPICS Input-Output Controller (IOC) was implemented into the embedded processor, using EPICS 3.15.4.

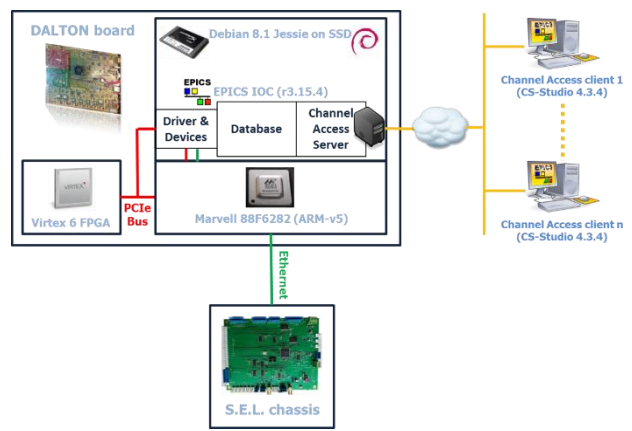

Figure 8: LLRF Control System (CS) hardware and software architecture.

The IOC was designed in order to communicate with two pieces of hardware: the DALTON board through a PCIe bus, and the SEL using an Ethernet client-server communication, as presented in Fig 8. Both interfaces were implemented as C libraries and C++ classes, integrated directly into the EPICS Driver layer, whose functions are then used by the EPICS Device layer (Process Variable processing). Once operational, it allowed to define the Process Variables (PVs) database that collects information from the hardware and the firmware and describes the LLRF. This pool of PVs was finally made available to remote client applications (OPIs) using the EPICS Channel Access (CA) facility, which led to a complete CC of the LLRF system. It was in addition enriched and upgraded during the tests campaigns to better fit the end-user operational needs.

The General User Interfaces (GUIs) realized with Control System Studio 4.3.4 (CSS), as shown in the example Fig 9.

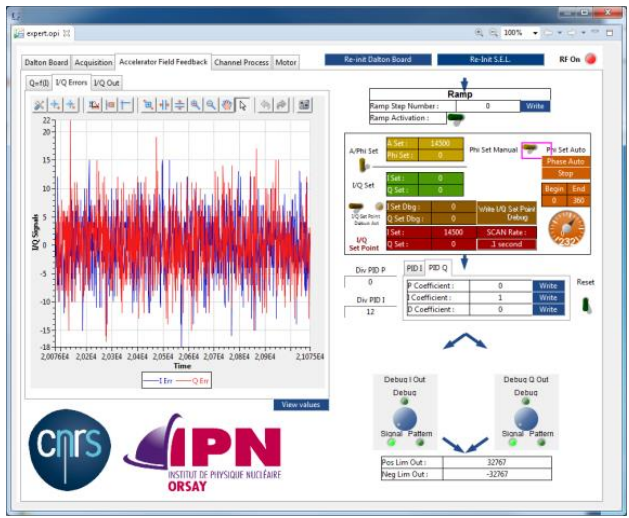

Figure 9: Example of an OPI of the MYRTE LLRF, designed with CSS

## TEST RESUTLS WITH RFQ

After a first RFQ conditioning realized by the SCK.CEN until 145KW, the first power tests campaign has been realized with the LLRF system. The Self excited loop has been used to increase to the nominal power, 110kW, and validate the LLRF RF power calibration. It transpired that the RFQ frequency variation was low allowing to start directly in Generator Driven (GDR) mode for the next time. Consequently the SEL mode will be used for the conditioning in the future.

A second step concerned the increase of the power from low power in the GDR mode with regulation. The first test has been limited in power at 30 kW in order to find the adapted PI parameters taking into account the gain variations amplifier under 70 kW. Let us remind that the operational set point is 110 kW with a usable range for regulating between 83KW and 145 kW.

The frequency tuning has been realized controlling the motorized plunger in manual mode via the CSS GUI.

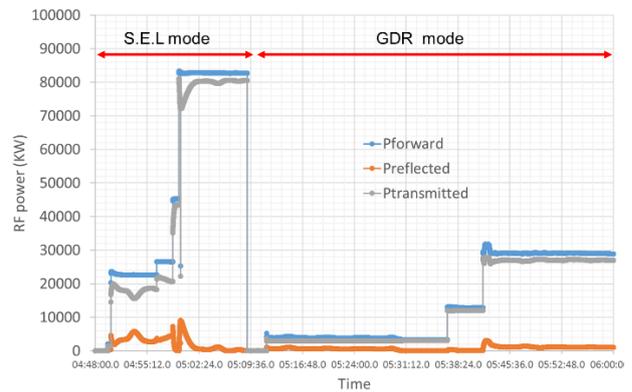

Figure 10: First tests in SEL mode and GDR mode without an automatic frequency tuning.

The second campaign's goal was to validate the automatic frequency tuning until the nominal set point, with in parallel the amplitude and phase regulation activated.

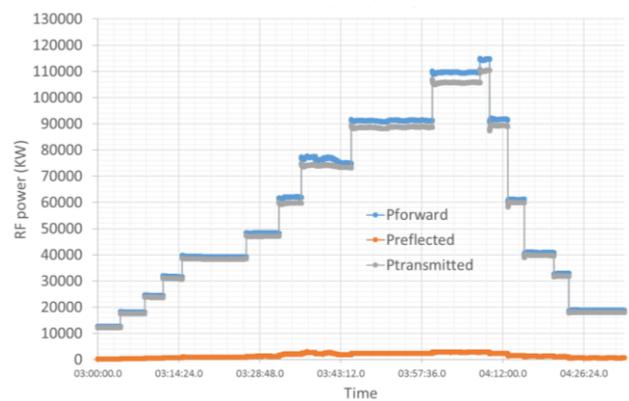

Figure 11: Tests in GDR mode with automatic frequency tuning and field regulation.

The first results show in Fig 12, an automatic follow-up of the frequency tuning during the RF power rise with some manual adjustments of the set point at 70kW and 110kW due to the temperature increase, for keeping the best return loss we can get. The last adjustment will be used for the future RF power rises. The 10 percent evolution of the plunger position in function of the RF power shows also the temperature impact.

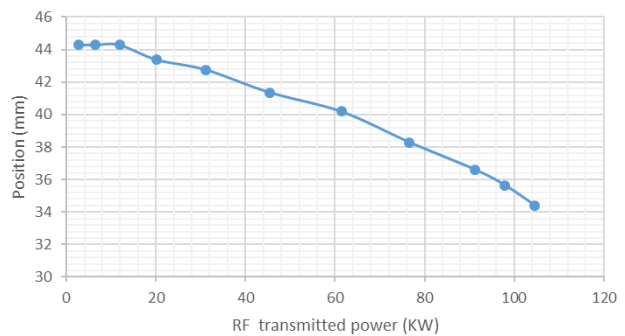

Figure 12 plunger position in function of the RF power

Concerning the amplitude and phase, the stability values were greatly improved thanks to the gain augmentation of the RF power amplifier with values better than 0.1% and 0.1°. But the risk was to find instabilities at high RF power as the measurements in open loop has shown low gain and phase margins, not precisely quantified for the moment. It's sure that it is not sufficient for a reliable and robust system, so two modifications are planned for reducing the latency and then to find a good trade-off concerning the margins versus stability performances.

Table 2: Latency.

| Components | Latency |
| --- | --- |
| FPGA (ACD + VHDL blocks) | 1.475 µs |
| SSPA (Measured with cables =0.4µS) | 0.100 µs |
| Cables (estimate : 5ns/m) | 0.300 µs |
| Downconverter (Measured) | 1.500 µs |
| Total Latency | 3.375 µs |

As shown in the table 2, the main latency is due to the downconverter and the signal processing performed by the FPGA. The goal is to reduce at least by a factor three the total latency in order to unsure a phase margin equal to 40 degrees and a response time inferior to the 3.8 µs corresponding to the filling time of the RFQ.

To achieve the above goal, the IF frequency must be increased in order to use IF filters with low group delay. An IF frequency equal to 20 MHz was chosen allowing to keep a Signal-to-Noise Ratio (SNR) superior to 70dB and multiply by two the clocks used for the sampling and the signal processing (80MHz and 160 MHz). Indeed, the FPGA clock is limited to 200MHz max. The characterization of the new down converter board is in progress.

## CONCLUSIONS

The in-house LLRF system designed by IPNO is validated with the RFQ (without beam) at the nominal RF power (110kW). However some improvements are in progress to increase margins of stabilities and facilitate the operator's job. Validation with the beam is planned for next year. In parallel, a LLRF mTCA (Micro Telecommunication Computing Architecture) based development [6] is on-going within a collaboration contract between SCK•CEN and IN2P3, benefiting from MYRTE developments and tests.

## ACKNOWLEDGEMENTS

The presented work was made possible by the collaboration with SCK•CEN, IAP Frankfurt, NTG and COSYLAB. Special thanks to SCK•CEN and UCL teams at Louvain-La-Neuve for their involvement during installation and tests of the LLRF system.

## REFERENCES

[1] J.-L. Biarrotte et al., "Design of the MYRRHA 17-600 MeV superconducting linac", SRF2013, Paris, France.

[2] H. Podlech et al., "The Proton Linac for the MYRRHA Project", TCADS-3, Mito, Japan, 2016.

[3] D. Mäder et al., "Status and Development of the MYRRHA Injector", IPAC'18, Vancouver, Canada.

[4] F. Bouly et al., "Commissioning of the MYRRHA low energy beam transport line and space charge compensation experiments", *IPAC'17*, Copenhagen, Denmark.

[5] F. Pompon, "Commissioning status of the RFQ Solid-State Amplifier", presentation during the 7th WP2 MYRTE meeting, CERN, 2018.

[6] W. Sarlin, Y. Gargouri, C. Joly et al., "A µTCA-based Low Level RF system prototype for MYRRHA 100 MeV project", LRF Workshop 2019, Chicago, USA.